\documentclass[12pt]{article}
\usepackage[utf8]{inputenc}
\usepackage{lipsum}
\usepackage{hyperref}
\usepackage{enumitem}
\usepackage{float}
\usepackage{graphicx}
\usepackage{caption} 
\usepackage{calc}
\usepackage{mathrsfs}
\usepackage{fancyhdr}
\usepackage{tcolorbox}
\usepackage{hyperref}
\usepackage{wrapfig}
\usepackage{graphicx} 
\usepackage{amssymb}
\usepackage{amsmath}
\usepackage{multicol}
\usepackage{braket}
\usepackage{abstract}
\usepackage{multicol}
\usepackage{xcolor}
\usepackage{appendix}
\usepackage{cite}
\title{\normalsize\bf Near-Horizon BMS Symmetry and  Implications on Black Hole Entropy}
\author{\normalsize {\sc Nihar Ranjan Ghosh\footnote{\tt g.nihar@iitg.ac.in}\ \ and Malay K. Nandy\footnote{\tt mknandy@iitg.ac.in {\rm (Corresponding Author)}}}\\
\normalsize \em Department of Physics, Indian Institute of Technology Guwahati\\
\normalsize \em Guwahati 781 039, India}
\date{\small (August 27, 2026)}

\begin{document}

\maketitle

\begin{abstract}
Thermodynamic significance of near-horizon symmetries remains an important open question in black hole physics, particularly in the context of black hole evaporation and information recovery. In this work, we investigate the role of horizon-adapted Bondi-van der Burg-Metzner-Sachs (BMS)-like supertranslations in the thermodynamic description of a dynamical Schwarzschild black hole. Working in a near-horizon Rindler coordinate system, we construct a class of diffeomorphisms that preserve the horizon structure and promote the associated supertranslation parameter to a Goldstone-like mode arising from the breaking of horizon symmetry. By expanding the Einstein-Hilbert action around the background geometry, we obtain the effective action for the Goldstone mode and identify the corresponding conserved horizon charge from the surface contribution of the action. The relevant horizon is defined at the future outer trapping horizon, while the surface gravity is computed using the Kodama-vector construction appropriate for dynamical spacetimes. We show that the horizon supertranslation mode contributes non-trivially to the surface gravity and modifies the thermodynamic description of the black hole beyond the stationary limit. Using the associated Noether charge, we derive the entropy of the horizon-BMS transformed geometry and find that the Bekenstein-Hawking area law is recovered at leading order, while subleading corrections depend explicitly on the supertranslation sector and the dynamical evolution of the black hole.
\end{abstract}

\maketitle
\tableofcontents
\section{Introduction}
Black holes occupy a unique place in gravitational physics because they unite geometry, thermodynamics, and quantum theory in a single setting. The discovery by Bekenstein and Hawking that black holes possess thermodynamic properties, with the entropy proportional to the horizon area, provided one of the earliest and most striking hints that gravity may admit a microscopic statistical description \cite{PhysRevD.7.2333, hawking1975particle}. Later on Hawking's finding of black hole evaporation \cite{PhysRevD.14.2460} implies the destruction of information when a black hole is formed and subsequently evaporates \cite{Hawking:1975vcx, hawking1974black}. This destruction of information leads to non-unitary evolution from a pure state of the black hole to a mixed state, which is forbidden in quantum mechanics. Even though several attempts have been made to resolve this paradox both in classical and quantum gravity, the first possible resolution came through the work of Hawking et.al. \cite{PhysRevLett.116.231301}. They have shown that the black holes carry an infinite number of soft charges which can, in principle carry the information along with them. These infinite number of soft charges (low energetic) are the symmetry charges corresponding to the asymptotic symmetry of all asymptotic flat spacetimes, first discovered by Bondi-van der Burg-Metzner-Sachs \cite{bondi1962gravitational,sachs1962gravitational}, commonly known as BMS symmetry  group.

The BMS symmetry group represents the infinite dimensional symmetry group of all asymptotically flat spacetimes, by enlarging the finite-dimensional Poincar\'e group of flat spacetime by an infinite number of generators known as {\em supertranslations}. These transformations can also be interpreted by angle-dependent translations near null infinity, and they significantly enrich the phase space structure of general relativity by relating spacetime configurations that are otherwise distinguishable under Poincar\'e symmetry \cite{Ghosh:2025dkn}. Moreover, in a recent study, this infinite number of symmetry generators was found to be constrained via classical energy conditions \cite{Ghosh:2026cno}. Similar development occurred in the context of asymptotically anti-de Sitter (AdS) spacetimes, notably through the work of Brown and Henneaux \cite{brown1986central}. While the focus of this work remains on asymptotically flat spacetimes, the structural similarities in both settings highlight the universality and physical relevance of asymptotic symmetry analysis \cite{CADONI1999165, hotta1998asymptoticisometrydimensionalantide, henneaux1985asymptotically, Comp_re_2016}. This asymptotic BMS symmetry was found to be deeply connected with two other apparently uncorrelated developments known as soft theorems and memory effects. Together, these three seemingly different topics constitute the three corners of the famous {\em infrared triangle} \cite{strominger2018lecturesinfraredstructuregravity}.

The gravitational memory effect, one of the three corners of the infrared triangle, is the permanent displacement effect produced by gravitational radiation, first identified in the linearized regime \cite{Zeldovich:1974gvh} and subsequently extended to the nonlinear case \cite{Christodoulou, Braginsky:1985vlg, Braginsky:1987kwo, PhysRevD.44.R2945, PhysRevD.46.4304, PhysRevD.45.520, Favata:2010zu, Tolish:2014bka, Tolish:2014oda, Winicour:2014ska}. For a more detailed study, the reader is invited to the Refs.~\cite{Strominger:2014pwa, strominger2018lecturesinfraredstructuregravity, PhysRevD.92.084057, Flanagan:2015pxa, Pasterski:2015tva, Pasterski:2015zua, PhysRevD.98.064032, Compere:2016jwb, Bieri:2013hqa, susskind2019electromagneticmemory, PhysRevLett.116.091101, PhysRevD.102.044041}. The third corner of the infrared triangle is formed by soft theorems, originating from infrared studies in quantum electrodynamics \cite{PhysRev.52.54, Low:1954kd, Low:1958sn, Gell-Mann:1954wra, Yennie:1961ad}. In gravity, Weinberg established the universal soft-graviton theorem \cite{Weinberg:1965nx}, later shown to be equivalent to Ward identities of asymptotic symmetries \cite{Pasterski:2015tva}, thereby providing the quantum realization of BMS symmetries and the foundation of the soft-hair paradigm \cite{PhysRevLett.116.231301, Hawking:2016sgy, haco2018black, PhysRevD.96.084032, Chu_2018, PhysRevD.108.044034, strominger2017blackholeinformationrevisited, haco2019kerrnewmanblackholeentropy, PhysRevD.103.126020, Donnelly:2014fua}. More precisely, the quantum Ward identities associated with the supertranslation and superrotation symmetry of the gravitational $S$-matrix are equivalent to the leading and subleading orders of the soft graviton theorem \cite{Strominger:2013jfa, He:2014laa, Campiglia:2014yka, Campiglia:2015yka}.

In spite of these developments in black hole thermodynamics and BMS symmetry, some important fundamental implications of the BMS group and the resulting thermodynamic properties remain unexplored. 
\begin{enumerate}
    \item First, even though the BMS group has been extensively explored at null infinity in stationary black hole settings, its implications for dynamical horizons needs attention. In this context, it may be noted that a complementary perspective interprets the soft degrees of freedom as edge modes or Goldstone excitations associated with large gauge transformations that act nontrivially at spacetime boundaries \cite{Donnelly:2014fua, Donnelly:2015hxa, Harlow:2015lma, Harlow:2016vwg, Maldacena:2016upp}. The black hole horizon may therefore be viewed as a boundary supporting Goldstone modes generated by spontaneously broken BMS symmetries \cite{Eling:2016xlx, averin2016gravitational, mpla, PhysRevLett.116.231301, hawking2017superrotation, MAITRA2022136825}.
    \item Second, although the existence of such modes has been widely discussed, their precise thermodynamic significance remains unclear. In particular, a concrete understanding of how horizon Goldstone modes contribute to black hole entropy is still lacking, especially in dynamical spacetimes. \item Third, although the Bekenstein–Hawking area law $S=A/(4G)$ is well-understood in the context of quantum field theory in curved background in the stationary case, extending this idea to truly dynamical black holes with broken spherical symmetry is subtle. In the absence of a timelike Killing vector, concepts such as surface gravity and entropy become local properties of evolving horizons. This motivates the study of quasi-local horizons (e.g. future outer trapping horizons) and their thermodynamics.  
    \item And finally, an important open question is whether the Noether charges associated with these supertranslation modes can account for the microscopic degrees of freedom responsible for black-hole entropy. More generally, it remains unclear how the infinite-dimensional horizon symmetry structure is encoded in the thermodynamic properties of evolving black holes.
\end{enumerate}

Motivated by these issues, we consider in this work a Vaidya-Schwarzschild black hole and study the Goldstone modes generated by horizon supertranslations. Treating the event horizon as a boundary on which BMS-like diffeomorphisms act nontrivially \cite{koga, Iofa}, we construct the corresponding Goldstone action and investigate the associated Noether charges. Since black-hole entropy is expected to originate from boundary contributions to the gravitational action \cite{Hawking-path-1, Hawking-path-2}, our analysis focuses on the surface sector of the Goldstone action rather than on the bulk action which governs the dynamical evolution equations for these modes. Moreover, to get a dynamical picture of the scenario, one needs to solve the dynamical equation (Euler-Lagrange equations for higher order derivative theory) for these Goldstone modes which are coupled to the background spacetime through Einstein's equations where the energy momentum tensor corresponds to these Goldstone modes.

The model considered here can be understood as follows. We consider a Vaidya- \linebreak Schwarzschild black hole which is supertranslated by a BMS generating vector field, and due to the non-trivial action of the vector field, the geometry near the black hole horizon changes and the spherical symmetry breaks down. As a result of the spontaneous symmetry breaking of supertranslation invariance, Goldstone modes are generated that live on the horizon and evolve both in time and along the angular directions. These Goldstone bosons constitute the supertranslation parameter of the symmetry transformation. Through the association of the Vaidya geometry with the Goldstone modes we can identify them with Hawking radiation. By calculating the conserved Noether charge on the horizon corresponding to the boson action,  we find that it reproduces the standard $S=A/(4G)$ entropy law at the leading order, with the subleading contributions originating from the Goldstone mode dynamics. Even though the primary objective here is to examine whether these horizon degrees of freedom reproduce the expected entropy relation and thereby clarify the thermodynamic role of horizon BMS symmetry in dynamical black-hole backgrounds, the stage set here acts as an initial basis to check the behaviour of the coupled black hole-Goldstone mode dynamics.

The rest of the paper is organized as follows. In Section \ref{sec. Gauge}, we introduce the gauge conditions and the metric perturbation relevant to the Vaidya-Schwarzschild geometry. Section \ref{sec. action} is devoted to deriving the effective action, conserved current, and Noether charge associated with the Goldstone mode. In Section \ref{sec. trapped horizon} we discuss the trapping horizon and its role in defining surface gravity for the dyanmic spacetime. Section \ref{sec. surface gravity} presents the analysis of the surface gravity employing the Kodama vector, while Section \ref{sec. entropy} establishes a connection between charge and entropy for the supertranslated Vaidya-Schwarzschild geometry. Finally, in Section \ref{sec. discussion}, we summarize our main results and present concluding remarks.

\section{Gauge Conditions and Metric Perturbation}{\label{sec. Gauge}}
As entropy is an intrinsic property of a black hole, a proper study would require the analysis on its horizon rather than at the asymptotic $r\to\infty$ limit. Therefore, in this study, to calculate the entropy, we consider the class of BMS-type horizon adapted symmetry, which preserves the horizon structure.  In doing so, we choose a coordinate system which is particularly suited for near horizon analysis, namely the Rindler coordinate system. The metric of a static Schwarzschild black hole ($m$=constant), expressed in a near horizon coordinate system is given by  
\begin{equation}
    \label{rindler coordinate}
    ds^2=-\frac{\rho^2}{16m^2}dv^2+\frac{\rho}{2m}dvd\rho+4m^2d\Omega_2^2~,
\end{equation}
with $d\Omega_2^2=d\theta^2+\sin^2\theta d\phi^2$ the metric of the $2$-sphere. The horizon of the geometry is located at $\rho=0$. Now defining a new coordinate as $r=\rho^2/(8m)$, we have from equation \ref{rindler coordinate}
\begin{equation}
    \label{metric}
    ds^2\Bigg|_{g_{ab}}=-\frac{r}{2m}dv^2+2dvdr+4m^2d\Omega_2^2~.
\end{equation}

Although near-horizon geometry in stationary black-hole settings has been explored in a number of works, its applications for dynamical horizons remain much less understood. Moreover, to associate the supertranslation parameter (Goldstone modes) with the Hawking radiation, one needs a dynamic spacetime. The Vaidya–Schwarzschild geometry provides one of the simplest and most useful models of a non-stationary black hole, describing evaporation through a time-dependent mass function. This geometry is obtained with the transformation $m\to m(v)$. Thus, the near horizon metric for a Vaidya Schwarzschild black hole, can be written as
\begin{equation}
    \label{original metric}
    ds^2\Bigg|_{g_{ab}}=-\frac{r}{2m(v)}dv^2+2dvdr+4m(v)^2d\Omega_2^2~
\end{equation}
Interestingly, the volume element of the $v$-$r$ constant hypersurface is now time dependent, which is expected as the metric represents the near horizon $(r=0)$ geometry of a Schwarzschild Vaidya black hole, and as the black hole evaporates or accumulates mass its horizon radius changes and so does the volume element on the horizon.

Now as mentioned earlier, we will choose a particular set of gauge conditions which preserves the near horizon structure of the black hole in \ref{original metric}. One of such appropriate gauge choices are as follows: 
\begin{equation}
    \label{Lie-eta}
    \mathcal{L}_{\eta}g_{rr}=\mathcal{L}_{\eta}g_{rv}=\mathcal{L}_{\eta}g_{Ar}=0~~.
\end{equation}
The vector field $\eta$, which satisfies the above gauge conditions in equation \ref{Lie-eta} is given by 
\begin{equation}
    \label{eta}
    \eta=F\partial_v-r\partial_vF\partial_r-\frac{r}{4m^2 \gamma_{AA}}\partial_A~~,
\end{equation}
where $A=\theta,\phi$, with $\gamma_{AA}$ the $AA^{th}$ matrix component of the $2$-sphere metric. The arbitrary function $F(v,\theta,\phi)$ is known as the supertranslation parameter. 

The Goldstone theorem states that whenever a continuous global symmetry is spontaneously broken, there occurs a massless excitation about the spontaneously broken vacuum, known as Goldstone boson. In the context of BMS group, the horizon supertranslation symmetry is spontaneously broken due to the action of the vector field $\eta$, so that $F(v,\theta,\phi)$ appears as a Goldstone mode living on the modified horizon. Although $F$ originates as a gauge parameter, it parametrizes inequivalent near-horizon geometries (differing by soft hair) and can carry physical information.

In contrast to the asymptotic $r\to\infty$ BMS analysis, where the supertranslation parameter $F$ is only a function of the angular parameters, here in our analysis the gauge conditions are chosen so that  the supertranslation is also a function of the time coordinate $v$, which further enables us to treat it as the Goldstone boson of the spontaneously broken symmetry and to find its evolution in time, rather than on a constant $v$-$r$ hypersurface. This is essential if one hopes to connect horizon symmetries to dynamical black holes with $m=m(v)$.

As will be shown in the next part, under the action of the diffeomorphism generating vector field \ref{eta}, the underlying gravitational field $g_{ab}$ will be modified. This can be thought of as similar to the transformation which breaks the U$(1)$ symmetry of a scalar field $\Phi$. The modification in the geometry is parametrized by the supertranslation parameter $F(v,\theta,\phi)$. Consequently, the macroscopic parameters of the original metric will be modified and therefore with the analogy of U$(1)$ symmetry breaking, this can be regarded as breaking of the spherical symmetry of the horizon. Therefore, we can promote the parameter $F$ as Goldstone mode. Moreover, just like the action of the U(1) Goldstone mode is obtained from the scalar field Lagrangian, here also we shall obtain the corresponding action for  $F$ from the Einstein–Hilbert action.

Now because of the action of this vector field $\eta$ given by \ref{eta}, the new modified metric is
\begin{equation}
    \label{new metric}
    \begin{split}
        ds^2\Bigg|_{\bar{g}_{ab}}&=g_{ab}+\mathcal{L}_{\eta}g_{ab}\\
        &=r\left(-\frac{1}{2m}+\mathcal{F}_1\right)dv^2+2dvdr+r\partial_A\mathcal{F}_2dvdx^A+\left(4m^2+8mm'F-2r\partial_\theta^2F\right)d\theta^2\\
        &-2r(\partial_\theta\partial_\phi F-\cot\theta\partial_\phi F)d\theta d\phi+\left[(4m^2+8mm'F)\sin^2\theta-2r(\sin\theta\cos\theta\partial_\theta F+\partial_\phi^2F)  \right]d\phi^2~,
    \end{split}
\end{equation}
with $m=m(v)$, $\mathcal{F}_1=\frac{m'}{2m^2}-\frac{1}{2m}\partial_v F-2\partial_v^2F$ and $\mathcal{F}_2=\frac{2m'}{m}F-\frac{1}{2m}F-2\partial_vF$. It is important to note that because the original unperturbed metric was considered to be Vaidya type, the near horizon BMS transformation acts differently compared to the cases when one considers the background to be static. Therefore, the metric perturbation $h_{ab}$, generated by the action of the diffeomorphism generating vector field $\eta$ and parametrized by the supertranslation $F(v,\theta,\phi)$ is given by $h_{ab}=\mathcal{L}_{\eta}g_{ab}$, such that
\begin{equation}
    \label{hab}
h_{ab}=\begin{bmatrix}
r\mathcal{F}_1 & 0 & \frac{r}{2}\partial_\theta\mathcal{F}_2 & \frac{r}{2}\partial_\phi\mathcal{F}_2  \\
0 & 0 & 0 & 0  \\
\frac{r}{2}\partial_\theta\mathcal{F}_2 & 0 & \bar{g}_{\theta\theta}-4m^2 &  \bar{g}_{\theta\phi} \\
\frac{r}{2}\partial_\phi\mathcal{F}_2 & 0 & \bar{g}_{\theta\phi} & \bar{g}_{\phi\phi}-4m^2\sin^2\theta\\
\end{bmatrix}
\end{equation}
where $\bar{g}_{AB}$ can be found from equation \ref{new metric}. It is important to note that the action of the vector field $\eta$ breaks the global SO$(3)$ symmetry, as is clear from equation \ref{new metric}.

\section{Action for the Goldstone Mode and Conserved Charge}{\label{sec. action}}
As mentioned earlier, our analysis is based on the Einstein-Hilbert action expanded about the unperturbed Vaidya-Schwarzschild background. The perturbation induced by the horizon supertranslation generates an effective action for the corresponding Goldstone mode, which splits naturally into bulk and boundary contributions. Now, the Einstein-Hilbert action, written in terms of the new metric $\bar{g}_{ab} $ and new Ricci scalar $\bar{R} $ derived from the metric $\bar{g}_{ab} $, is 
\begin{equation}
    \label{EH action}
    S_E=\frac{1}{16\pi G}\int d^4x\sqrt{-\bar{g}}~\bar{R}~.
\end{equation}
If we consider the metric components as $\bar{g}_{ab}=g_{ab}+h_{ab}$, with $h_{ab}=\mathcal{L}_{\eta}g_{ab}$ as small fluctuations in the background metric $g_{ab}$, then to find the action corresponding to the Goldstone mode $F$, we can Taylor series expand equation \ref{EH action} as 
\begin{equation}
    \label{taylor}
    S_E\Bigg|_{\bar{g}_{ab}}=S_E[g_{ab}]+h_{ab}\left(\frac{\delta S_E}{\delta \bar{g}_{ab}} \right)\Bigg|_{g_{ab}}+h_{ab}h_{cd}\left(\frac{\delta^2 S_E}{\delta \bar{g}_{ab}\delta \bar{g}_{cd}}  \right)\Bigg|_{g_{ab}}+\dots
\end{equation}
The first term on the right hand side does not contribute to the analysis. Moreover, the second term vanishes as the background metric $g_{ab}$ is a solution of the Einstein's equations of motion and hence the second term is proportional to the Einstein tensor. Non-trivial contribution comes from only the third term on the right hand side which governs fluctuations around the vacuum. This is analogous to Goldstone excitations in spontaneous symmetry breaking. Therefore, $F$ describes low-energy excitations of the horizon geometry. Thus, the third term is taken to be the action corresponding to the Goldstone mode $F$. The higher power terms in $h_{ab}$ are expected to contribute in the sub-leading order as we treat $h_{ab}=\mathcal{L}_{\eta}g_{ab}$ as small fluctuations in the background spacetime $g_{ab}$. 

As explained earlier, we shall concentrate on the surface part of the action and use its diffeomorphism invariance to construct the associated conserved current and antisymmetric Noether charge $2$-form. This charge will then be evaluated on the horizon and interpreted as the charge carried by the horizon-BMS sector in Section \ref{sec. entropy}.

Following the analysis and toolkit given in \cite{Ghosh:2026cno}, we find that the third term on the right hand side of equation \ref{taylor} can be written as a sum of the bulk  and surface contributions, $S_{\rm bulk}$ and $S_{\rm surface}$, with 
\begin{equation}
\label{surface term}
    \begin{split}
        S_{\rm surface}=\frac{1}{16\pi G}\int d^4x\sqrt{-g}\nabla_a\left[\frac{1}{2}h\nabla_b h^{ab}-\frac{1}{4}h\nabla^ah-h^{bc}\nabla_bh^a_c-h^{ab}\nabla_ch^c_b        \right]~,
    \end{split}
\end{equation}
which can be written as $S_{\rm surface}=\int d^4x \sqrt{-g}\nabla_aA^a$, where 
\begin{equation}
    \label{A term}
    A^a=\frac{1}{16\pi G}\left[\frac{1}{2}h\nabla_b h^{ab}-\frac{1}{4}h\nabla^ah-h^{bc}\nabla_bh^a_c-h^{ab}\nabla_ch^c_b        \right]~.
\end{equation}

Having obtained the surface term of the action for the Goldstone mode, by using its diffeomorphism invariance, we shall construct the conserved current and conserved charge for a general class of action. Consider a generic action $S$ that can be written as a total derivative of a vector field $B^a$, implying 
\begin{equation}
    \label{label 1}
    \sqrt{-g} \mathscr{L}=\sqrt{-g}\nabla_aB^a~.
\end{equation}
Now, under a diffeomorphism $\zeta$, the left hand side of equation \ref{label 1} changes as 
\begin{equation}
    \label{lhs}
    \delta_{\zeta}(\sqrt{-g} \mathscr{L})\equiv\mathcal{L}_{\zeta}(\sqrt{-g} \mathscr{L})=\sqrt{-g}\nabla_a(\mathscr{L}\zeta^a)~,
\end{equation}
whereas the right hand side changes as
\begin{equation}
    \label{rhs}
    \delta_{\zeta}(\sqrt{-g}\nabla_aB^a)=\sqrt{-g}\nabla_a\left[\nabla_b(B^a\zeta^b)-B^b\nabla_b\zeta^a    \right]~.
\end{equation}
Equating equations \ref{lhs} and \ref{rhs}, we get the current $J^a$ satisfying $\nabla_aJ^a=0$, with
\begin{equation}
    \label{current}
    J^a[\zeta]=\mathscr{L}\zeta^a-\nabla_b(B^a\zeta^b)+B^b\nabla_b\zeta^a~,
\end{equation}
which, upon using equation \ref{label 1},  gives the charge $2$-form $Q^{ab}$ satisfying $\nabla_a\nabla_bQ^{ab}=0$,
\begin{equation}
    \label{charge}
    Q^{ab}=(\zeta^aB^b-\zeta^bB^a)=2\zeta^{[a}B^{b]}~,
\end{equation}
where $\zeta^{[a}B^{b]}=\frac{1}{2}(\zeta^aB^b-\zeta^bB^a)$. The definition of charge $2$-form in equation \ref{charge} is valid for any Lagrangian that can be written as a total derivative and for any diffeomorphic vector field $\zeta$. Thus, we can identify $\zeta$ with $\eta$ in equation \ref{eta} and the vector $B^a$ with $A^a$ in equation \ref{A term}, to calculate the charge corresponding to the symmetry related to the diffeomorphic vector field $\eta$. Relating the action in \ref{surface term} with an antisymmetric tensor $Q^{ab}$ simplifies the rest of the analysis, as although the original action was in terms of an integral over the $4$D manifold, it can now be converted to an integral over a $2$D hypersurface. Moreover, the existence of a conserved charge associated with the horizon symmetry suggests that the corresponding soft sector contributes to the microscopic state count of the horizon. Since black hole entropy is expected to encode the number of accessible microscopic configurations, it is natural to investigate whether the soft horizon charges influence the thermodynamic entropy.

Now, for Einstein gravity with $g_{ab}$ of a stationary black hole and $\xi$ as a horizon generating Killing vector, we have
\begin{equation}
    \label{einstein charge}
    Q^{ab}=-\frac{1}{16\pi G}\nabla^{[a}\xi^{b]}~,
\end{equation}
so that, with $\kappa$ as surface gravity, the following integral
\begin{equation}
    \label{einstein entropy}
    \frac{2\pi}{\kappa}\int_{\mathscr{H}} d\Sigma_{ab}Q^{ab}=\frac{A}{4G}=\mathcal{S}
\end{equation}
with $\mathscr{H}$ as the horizon, gives the entropy of the black hole.

In this work, we generalize this definition to calculate the entropy of a dynamical horizon-BMS transformed Schwarzschild black hole. In doing so, the very first step would be to define the surface gravity $\kappa$ for a dynamic spacetime with broken spherical symmetry. However, the notion of surface gravity in such a setting needs a careful treatment. Since the spacetime is no longer stationary, the event horizon is not the most natural local entity for thermodynamic analysis. Instead, a quasi-local description based on a trapped surface and a future outer trapping horizon (FOTH) become more appropriate, since it allows one to formulate horizon dynamics and entropy in a way that is adapted to time-dependent geometries. In such settings, the notion of surface gravity also requires a more flexible definition; the Kodama-vector construction is particularly well-suited for this purpose because it reproduces the standard Killing result in the stationary limit while remaining applicable to dynamical spacetimes. As will be clear in the next section, the surface gravity is obtained particularly by performing the analysis on the horizon itself, and hence the horizon adapted BMS transformations considered here makes the analysis more robust compared to an asymptotic $r\to\infty$ BMS analysis.

\section{Trapped Horizon}{\label{sec. trapped horizon}}
In order to calculate the surface gravity, an important ingredient is to identify the location of the event horizon of the spacetime. The calculation of the surface gravity on this particular hypersurface will be considered in the next section. Once we identify the location of the event horizon, we can use Hawking's theorem which says that the event horizon of a stationary asymptotically flat spacetime is also a Killing horizon. On this Killing horizon, using the null Killing vector, we can associate a scalar called the surface gravity $\kappa$ which directly connects the Hawking temperature and entropy in black hole thermodynamics.

However, in the general case of dynamical black holes, there is no assurance of finding a Killing horizon, and hence in general one cannot invoke Hawking's theorem to find the surface gravity as there is no knowledge of the surface on which one should assign the surface gravity. A great deal of effort has been made to find a suitable local definition of horizons in dynamical black hole spacetimes in terms of {\em trapped surfaces}, which are spacelike $2$-surfaces where the expansion of the outgoing null rays normal to the surface vanishes \cite{Hayward, ashtekar1999isolated, ashtekar2002dynamical, ashtekar2004isolated, hayward2009dynamics, PhysRevLett.92.011102, faraoni2013evolving}. A trapped surface is an important concept in gravity since, under certain criterion, they lead to caustic formation. The cosmic censorship hypothesis suggests that there must be an event horizon, with the trapped surface located inside the black hole horizon. However, the dynamical spacetime is a highly non-trivial problem because in this case the horizon might not be a null hypersurface although it should still exhibit infinite redshift. In this work, we are going to follow this particular case in finding the position of the trapped surface on which we shall calculate the surface gravity.

We consider two bundles of null geodesics with tangent vectors $k^a$ and $l^a$ such that  $k^ak_a=l^al_a=0$ and $k^al_a=-1$. Then the induced metric of the hypersurface $\mathbb{S}$, orthogonal to both $k^a$ and $l^a$, can be written as
\begin{equation}
    \label{q}
    q_{ab}=\bar{g}_{ab}+l_ak_b+k_al_b~.
\end{equation}
Now, associated to these flows, characterized by $k^a$ and $l^a$, we have the expansion scalar $\Theta$, defined as $\Theta=q_{ab}\nabla^a n^b$, with $n^a=l^a,k^a$. Physically $\Theta$ measures the expansion rate of the surrounding radial null geodesics, with the radial null geodesics expanding (contracting) if $\Theta>0$ ($\Theta<0$). Now a spacelike closed  $2$-surface in $4$D has two independent normal directions, corresponding to the ingoing and outgoing null rays. Therefore, we can take $l^a$ and $k^a$ to be the tangent vectors of the bundles of outgoing and ingoing null geodesics to check the behaviour of the corresponding $\Theta$ to find the nature of the gravitational field surrounding $\mathbb{S}$. The trapped surface is identified as the surface for which $\Theta_l<0$ and $\Theta_k<0$ and a marginally trapped surface is the one for which $\Theta_k<0$ and $\Theta_l=0$, i.e. the outgoing null rays momentarily stop expanding. Although several proposals for defining black hole horizons exist, the future outer trapping horizon (FOTH) by Hayward is of particular importance \cite{Hayward}.

A future outer (marginally) trapping horizon is a smooth three-dimensional submanifold of spacetime which is foliated by closed space-like surfaces $\mathbb{S}_t$, $t\in\mathbf{R}$, with null normals $l^a, k^a$ constructed by ingoing and outgoing rays such that \cite{chu2018soft}
\begin{equation}
\label{trap}
    \Theta_l=0~~\text{(Marginally trapped)}~~\Theta_k<0~~\text{(Future type)}~~k^a\partial_a\Theta_l<0~~\text{(Outer type)}
\end{equation}
 
 The first condition in \ref{trap} specifies the location of the marginally trapped surface where nearby surrounding radial outgoing null geodesics are parallel. The second condition says that the trapping horizon is of future type, i.e. a black hole rather than a white hole. The third condition says a motion of $\mathbb{S}_t$ along $k^a$ makes it trapped, hence it is outer rather than inner type.
 
 Thus, with the above arguments we have the corresponding tangent vectors as
\begin{equation}
    \label{ka and la}
    k^a=\left[0,-1,0,0  \right]~~\text{and}~~l^a=\left[1,-\frac{r\alpha}{2}+\frac{r^2}{2}\bar{\gamma}^{AB}\beta_A\beta_B,-r\bar{\gamma}^{AB}\beta_B  \right]~,
\end{equation}
with $\alpha=-\frac{1}{2m}+\mathcal{F}_1$ and $\beta_A=\frac{1}{2}\partial_A\mathcal{F}_2$ where $\bar{\gamma}_{AB}$ is the metric element of the $2$-sphere corresponding to the modified metric $\bar{g}_{ab} $.
Therefore, the induced metric is given by
\begin{equation}
    \label{qab}
q_{ab}=\begin{bmatrix}
\frac{r^2}{\bar{g}_{\theta\theta}\bar{g}_{\phi\phi}-\bar{g}_{\theta\phi}^2}\left(\bar{g}_{\phi\phi}\beta_\theta^2+\bar{g}_{\theta\theta}\beta_\phi^2-2\bar{g}_{\theta\phi}\beta_\theta\beta_\phi  \right) & 0 & r\beta_\theta & r\beta_\phi  \\
0 & 0 & 0 & 0  \\
\frac{r}{2}\partial_\theta\mathcal{F}_2 & 0 & \bar{g}_{\theta\theta} &  \bar{g}_{\theta\phi} \\
\frac{r}{2}\partial_\phi\mathcal{F}_2 & 0 & \bar{g}_{\theta\phi} & \bar{g}_{\phi\phi}\\
\end{bmatrix}
\end{equation}
The expansion scalar $\Theta$ corresponding to the two null vectors at $r=0$ are 
\begin{equation}
    \label{expansion l}
    \Theta_l\Bigg|_{r=0}=\frac{2}{m(m+2Fm')}\left[F(m')^2+m\left\{Fm''+m'\left( 1+\partial_vF\right)\right\}  \right]
\end{equation}
and 
\begin{equation}
    \label{expansion k}
     \hspace{-4cm}\Theta_k\Bigg|_{r=0}=\frac{\csc^2\theta\partial_\phi^2F+\cot\theta\partial_\theta F+\partial_\theta^2F}{4m(m+2Fm')}
\end{equation}

By demanding that $r=0$ be the position of the FOTH, we impose $\Theta_l=0$ in equation \ref{expansion l}, leading to the generic form of $F(v,\theta,\phi)$ as
\begin{equation}
    \label{F cons}
    F(v,\theta,\phi)=\frac{1}{mm'}\Big[ f(\theta,\phi)-\frac{1}{2}m^2 \Big]~,
\end{equation}
with $f(\theta,\phi)$ some function of $(\theta,\phi)$, which can be further constrained using equation \ref{expansion k}.
Equation \ref{F cons} shows that the supertranslation parameter cannot be chosen arbitrarily once the FOTH condition is imposed. The horizon geometry and the supertranslation parameter become dynamically linked through the mass function $m(v)$. Consequently, the allowed horizon soft hair is constrained by the requirement that the transformed geometry continues to represent a physically acceptable black hole horizon.

\section{Surface Gravity}{\label{sec. surface gravity}}
Having obtained the position of the future of the outer trapping horizon (FOTH), the next step is to obtain the surface gravity for the supertranslated Schwarzschild black hole obtained in equation \ref{new metric}. For a stationary black hole, with $\xi$ as the horizon generating Killing vector, which vanishes on the bifurcation surface of the Killing vector, the surface gravity $\kappa$ is defined as
\begin{equation}
    \label{stationary kappa}
    \nabla_a\xi_b=\kappa\epsilon_{ab}~.
\end{equation}
 However, as there is no asymptotic timelike Killing vector for a dynamical black hole, the notion of surface gravity becomes more subtle and requires a delicate handling. In practice, depending on the local definition of horizons, one may define a notion of surface gravity on the trapped horizon \cite{Nielsen:2007ac}.

Nevertheless, there is quite a satisfactory and alternative definition of surface gravity for spherically symmetric geometry, as was proposed by Hayward \cite{Hayward:1997jp}. For any spherically symmetric metric, there exists a vector field called the Kodama vector $K$ that satisfies 
\begin{equation}
\label{kodama}
    \bar{\nabla}^a(\bar{G}^{ab}K^b)=0~~\text{and}~~\bar{\nabla}_aK^a=0~.
\end{equation}
For a spherically symmetric dynamic black hole, the Kodama vector gives a preferred time direction in the spacetime. Moreover, for an asymptotically flat spacetime, with an appropriate normalization, the Kodama vector coincides with the time translation Killing vector at spatial infinity. Therefore, this vector can safely be used in defining surface gravity for a dynamical geometry. 

Hayward has shown that on the FOTH, this vector field satisfies 
\begin{equation}
    \label{dynamic kappa}
    K^a\bar{\nabla}_{[b}K_{a]}=-\kappa K_b~,
\end{equation}
with $\kappa$ being the surface gravity of the dynamical geometry. The covariant derivative $\bar{\nabla}_a$ as well as the Einstein tensor $\bar{G}_{ab}$ in equation \ref{dynamic kappa} are computed with respect to the modified metric $\bar{g}_{ab} $ in equation \ref{new metric}. The Kodama vector becomes the time translation Killing vector and equation \ref{dynamic kappa} for surface gravity reduces to \ref{stationary kappa} in the stationary spacetime limit.

Following the definition in \ref{dynamic kappa} and using the ansatz that $K^a=\delta_v^a$, we find that equation \ref{kodama} is satisfied up to order $\mathcal{O}(F^2)$ and $\mathcal{O}((m')^2)$. Then, the surface gravity is obtained as
\begin{equation}
    \label{kappa}
    \kappa=\frac{1}{4m}+\frac{d}{dv}\left(\frac{1}{4m}  \right)+\frac{1}{4m}\partial_vF+\partial_v^2F
\end{equation}
As is well known that the surface gravity for  stationary Schwarzschild black hole is $\frac{1}{4m}$, the first two terms on the right side correspond to the surface gravity of a dynamic black hole and the last two terms correspond to the contribution coming from the BMS transformation. Therefore, temperature becomes sensitive to horizon symmetry and the Hawking radiation may therefore carry indirect information about the near horizon BMS transformation. Interestingly, for a stationary Schwarzschild black hole, equation \ref{kappa} reduces to the standard result $\kappa=1/(4m)$. It is also important to note that  equation \ref{kappa} depends on the angular coordinates through $F(\theta,\phi)$, along with the time coordinate $v$ for any non-trivial supertranslation. Therefore, the quantity obtained in equation \ref{kappa} can be considered to be the definition of {\em local} surface gravity. Consequently, the {\em effective surface gravity} is obtained by taking into account the contributions from all angular directions, implying $\kappa_{\rm eff}=\frac{\int \kappa d\theta d\phi}{\int d\theta d\phi} $.

\section{Charge and Entropy}\label{sec. entropy}
As mentioned earlier, we generalize equation \ref{einstein entropy} for our case of dynamic black hole to obtain its entropy after being supertranslated by the vector field $\eta$ from equation \ref{eta}. Therefore, using equation \ref{charge} with $A^a$ in equation \ref{A term}, we can integrate the charge $2$-form over the horizon $2$-surface to obtain the corresponding charge, related to the horizon symmetry. Thus, the charge on the horizon, at any constant time is given by
\begin{equation}
    \label{integral charge}
        Q=\int_{\mathscr{H}} d\Sigma_{ab}~Q^{ab}=\frac{m}{4G}\Big[1+\mathscr{F}(1/m)\Big]
\end{equation}
with $\mathscr{H}$ representing the horizon located at $r=0$, and $\mathscr{F}$ representing a polynomial function of $1/m$, which are sub-leading compared to the term $\frac{m}{4G}$.

Moreover, from the definition of $\kappa$ in equation \ref{kappa}, upon using equation \ref{F cons}, the effective surface gravity is found to be
\begin{equation}
    \label{final kappa}
   \begin{split}
        \kappa_{\rm eff}&=\frac{1}{8 m(v)}-\frac{m'(v)}{4 m(v)^2}+\frac{m(v) m^{(3)}(v)}{2 m'(v)^2}-\frac{m(v) m''(v)^2}{m'(v)^3}+\frac{m''(v)}{2 m'(v)}+\frac{m''(v)}{8 m'(v)^2}+\int f(\theta ,\phi ) d\theta d\phi\\
        &
        \times\Bigg[ \frac{2 m'(v) }{m(v)^3}-\frac{m^{(3)}(v)}{m(v) m'(v)^2}+\frac{2 m''(v)^2}{m(v) m'(v)^3}+\frac{m''(v)}{m(v)^2 m'(v)}-\frac{m''(v) }{4 m(v)^2 m'(v)^2}-\frac{1}{4 m(v)^3}  \Bigg]~,
   \end{split}
\end{equation}
that gives $\kappa_{\rm eff}=1/(8m)$ in the leading order. Unlike the stationary Schwarzschild case, the surface gravity now depends on both the evaporation history $m(v)$ and the horizon soft sector through $f(\theta,\phi)$. Therefore, the horizon temperature is not determined solely by the black hole mass but also by the state of the Goldstone modes. This suggests that the thermodynamic evolution of the black hole may retain information about horizon soft hair. Therefore, generalizing equation \ref{einstein entropy}, the entropy of the dynamical black hole is found to be
\begin{equation}
    \label{entropy}
    \begin{split}
        \mathcal{S}=\frac{A}{4G}~,
    \end{split}
\end{equation}
in the leading order of $m$, where $A$ is the surface area of the FOTH, and the sub-leading terms are proportional to $\frac{1}{m}$. It is important to note that the leading order term for entropy is independent of the choice of the supertranslation parameter $F(v,\theta,\phi)$.

\section{Discussion and Conclusion}{\label{sec. discussion}}
The possibility of relating horizon supertranslation to the evolution of black hole entropy provides the necessary motivation for understanding their thermodynamic significance. To explore this possibility, we investigated the role of horizon-adapted BMS-like transformations in the thermodynamic description of Schwarzschild black holes. The analysis was performed in a near-horizon Rindler coordinate system, where the horizon behaves as a null boundary, for stationary black hole, admitting non-trivial action of bulk diffeomorphisms. In this framework, the supertranslation parameter naturally appears as a Goldstone-like mode associated with the breaking of the horizon symmetry.

Unlike the standard asymptotic BMS construction, the supertranslation parameter considered here depends on the advanced time coordinate in addition to the angular coordinates. This choice is compatible with the near-horizon gauge conditions and allows the formalism to incorporate dynamical black hole geometries. The action of the corresponding diffeomorphism modifies the near-horizon metric while preserving the horizon structure, thereby generating non-trivial metric perturbations parametrized by the supertranslation parameter.

Starting from the Einstein--Hilbert action, we expanded the theory around the background geometry and identified the quadratic contribution in the perturbation as the effective action associated with the Goldstone mode. The resulting action separates into bulk and surface contributions. Since the thermodynamic properties of black holes are expected to be encoded in boundary terms, the analysis was restricted to the surface part of the action. Using the diffeomorphism invariance of the surface term, we constructed the associated conserved current and antisymmetric charge $2$-form, which in turn provided the horizon charge corresponding to the BMS-like symmetry.

To define the entropy in the dynamical setting, the relevant horizon was identified through the trapped-surface formalism. In particular, the future outer trapping horizon (FOTH) was obtained by imposing the vanishing of the outgoing null expansion together with the standard future and outer conditions. This procedure constrained the allowed form of the supertranslation parameter and provided a local definition of the horizon appropriate for dynamical, non-spherical geometries.

The surface gravity was then computed using the Kodama vector construction. This definition is well suited to dynamical spherically symmetric spacetimes and reduces to the usual Killing surface gravity in the stationary limit. For the supertranslated Schwarzschild geometry, the resulting surface gravity receives contributions both from the time dependence of the mass function and from the horizon supertranslation mode. The leading contribution reproduces the Schwarzschild value, while the remaining terms encode corrections induced by the dynamical and angular-dependent structure of the horizon.

Using the Noether charge associated with the surface action, we generalized the standard entropy relation to the horizon-BMS transformed geometry. The entropy was found to reproduce the Bekenstein-Hawking area law at leading order with subleading corrections depending on the supertranslation sector and derivatives of the mass function. Therefore, although the leading thermodynamic behaviour remains unchanged, the horizon soft sector contributes non-trivially beyond leading order. The corrections arise through the modified surface gravity and the associated horizon charge, indicating that the near-horizon symmetry structure influences the subleading thermodynamic properties of the black hole. Furthermore, even without addressing the full dynamics of the coupled Goldstone-gravity system, the present analysis reveals several non-trivial connections between near-horizon symmetries, conserved charges, and thermodynamic observables.

A notable feature of the present framework is that the horizon supertranslation mode enters simultaneously in the geometric, thermodynamic, and symmetry sectors of the theory. The same degree of freedom modifies the horizon geometry, contributes to the conserved charge, and affects the entropy through the surface gravity. This interrelation suggests that horizon soft modes are not merely gauge artifacts but may carry physically relevant information about the dynamical state of the black hole. Taken together, these results suggest that horizon-adapted BMS symmetries may provide a useful framework for understanding the microscopic origin of subleading corrections to black hole thermodynamics. More broadly, the present framework opens a promising avenue for exploring the interplay between horizon symmetries, dynamical horizons, and black hole thermodynamics.


\section*{Acknowledgement} 
Nihar Ranjan Ghosh is supported through a Research Fellowship from the Ministry of
Human Resource Development (MHRD), Government of India.


\end{document}